\documentclass[aps,prb,twocolumn,showpacs,superscriptaddress,amssymb]{revtex4}
\usepackage{graphicx}
\usepackage{amsmath}
\usepackage{hyperref}
\usepackage{natbib}
\usepackage{bm}
\begin{document}

\title{Time-optimal rotation of a spin 1/2: application to the NV center spin in diamond}

\author{Zhi-Hui Wang}
\author{V. V. Dobrovitski}
\affiliation{Ames Laboratory, Iowa State University, Ames IA 50011, USA}

\date{\today}
\begin{abstract}
We study the applicability of the time optimal bang-bang control designed
for spin-1/2 [U. Boscain and P. Mason, J. Math. Phys. {\bf 47}, 062101 (2006)]
to the rotation of the electron spin of a nitrogen-vacancy (NV) center in diamond.
The spin of the NV center is a three-level system, with two levels forming a
relevant qubit subspace where the time-varying
magnetic control field performs rotation, and the third level being idle.
We find that the bang-bang control protocol
decreases the rotation time by 20--25\% in comparison with the
traditional oscillating sinusoidal driving.
We also find that for most values of the bias field the leakage to the
idle level is very small, so that the NV center is a suitable testbed for
experimental study of the time-optimal protocols. For some special
values of the bias field, however, the unwanted leakage is greatly
increased. We demonstrate that this is caused by the resonance with
higher-order Fourier harmonics of the bang-bang driving field.
\end{abstract}
\maketitle

\section{Introduction}

Studying the dynamics of quantum spins helps to better understand many
fundamental issues of quantum physics, and is important for
applications in quantum information processing, \cite{NielsenChuang}
coherent spintronics, \cite{spintronics} and high-precision metrology. \cite{Jones09}
Recently, spins of the nitrogen-vacancy (NV) centers in diamond have emerged as a
particularly strong candidate for long-range quantum communication,
\cite{Childress06PRL} scalable quantum computation,\cite{JelezkoGate04,Dutt07} and
high-sensitivity magnetometry with nanoscale resolution.
\cite{Taylor08,Balasubramanian08,deLangeMagnetometry10}
NV centers also present an excellent platform for exploring such fundamental problems
as dynamics of quantum spins coupled to their environment,\cite{Childress06,Hanson08,Neumann10,deLange10}
quantum control and dynamical decoupling of single solid-state spins.\cite{Fuchs09,deLange10,Ryan10,Naydenov10}
This system has become very popular due to a favorable combination of properties:
a NV electronic spin has long coherence time, \cite{Childress06,Gaebel06,Hanson08,Balasubramanian09}
a single NV center can be optically initialized and read out, \cite{Gruber97,Jelezko02}
and can be manipulated both optically\cite{Santori06} and magnetically. \cite{Jelezko04}

In this paper we consider fast manipulation of the spin
of a single NV center by externally applied magnetic field,
which is important for many prospective applications.
The magnetic field can be quickly varied in time with modern electronics, but its
magnitude is always limited, so
the problem of time optimal spin control naturally arises:
how to shape the time profile of the control field
so that the desired spin evolution is accomplished within the shortest time.
Time-optimal control is a well developed science nowadays,\cite{DomenicoBook}
and the optimal control of quantum spins has been extensively studied in various
contexts, from general mathematical considerations of
the optimal control of two-level systems,\cite{DAlessandro01,Khaneja01,Boscain02,Boscain06,Carlini06,Sugny07}
to numerous applications, e.g.\ in the areas of magnetic resonance,\cite{Khaneja01,KhanejaGRAPE}
and quantum information processing.\cite{Palao02,WeninPotz06,RoloffPotz09,Fazio09,HsiehLidar10}
In most cases, when the system under consideration is sufficiently complex,
one has to use numerical methods to find the optimal control.\cite{Palao03,KhanejaGRAPE}
There are, however, rare cases where the exact solution is known,
and its optimality is rigorously proven. \cite{Boscain02,Boscain06}
It is of interest to see whether NV centers in diamond may
present a suitable system for experimental test of such fundamental results of the control theory.
Besides basic interests, such protocols can speed up control of NV spins in future
applications.

We consider the exact solution for time-optimal control of a spin 1/2,
placed in a static magnetic field along the $z$-axis, with the
corresponding energy splitting $2E$ between the states $|\uparrow\rangle$ and $|\downarrow\rangle$.
As in standard resonance experiments,\cite{Slichter}
a driving field $B_x(t)$ of limited amplitude, $|B_x(t)|\le M$, can be applied along the $x$-axis
to rotate the spin from $|\downarrow\rangle$ to $|\uparrow\rangle$.
The traditional solution is to use driving which oscillates with the spin's Larmor frequency $2E/\hbar$:
for small driving, $M\ll E$, within the rotating wave approximation (RWA), it
implements the desired rotation. However, this approach is not optimal; also, at strong driving the RWA breaks down,
making the spin dynamics intricate.\cite{Fuchs09,Ashhab07,Boyd00}
It has been proven \cite{Boscain06} that for a single-axis driving
the bang-bang control is optimal, with the driving field $B_x(t)$
switching between its extremal values.$\pm M$

Here we investigate whether this exact solution
is applicable to the spin of a NV center.
A NV center has spin $S=1$ with the anisotropy splitting $D=2.88$~GHz
between the levels $m_S=\pm 1$ and $m_S=0$.
Static external field applied along
the anisotropy axis further splits the levels $m_S=+1$ and $m_S=-1$.
In most applications, the states $m_S=0$ and $m_S=-1$
form the relevant two-level system (qubit, or a pseudo-spin 1/2)
while the third level $m_S=1$ remains idle.
However, when the optimal control designed for a spin-1/2 is
applied to the NV spin $S=1$, the idle third level can become occupied.
Protocols minimizing the leakage to higher levels 
have been developed before, using numerical tools, for
different physical systems and situations.\cite{Steffen03,Wilhelm09,Tian00}
But numerical optimization does not guarantee global optimality, and is model-dependent.
Thus, it is interesting to consider the opposite approach,
and study applicability of the known, rigorously optimal spin-1/2 protocol,
to a NV spin rotation (although such protocols could be sub-optimal when the full 3-level
system is considered).
We show that for most regions of the relevant parameter space,
the leakage to the idle level is negligible, so that
NV center does present a good testbed for the spin-1/2 protocol.
This also means that the known time optimal
protocol can be safely used to manipulate the NV center's (pseudo)spin,
and can noticeably, by $\sim 25$\%, speed up its rotation compared to RWA.
Moreover, we investigate the cases where the known solution
for spin 1/2 is not applicable, and explain the reasons for the
increased leakage to the idle NV spin state.

The rest of the paper is organized as follows. In Sec.~II, we describe
in more detail the spin Hamiltonian of the NV center, the time optimal
spin-1/2 protocol, and our results. Discussion and conclusions are given
in Sec.~III.

\section{Results: bang-bang control of the NV center and its performance}
\begin{figure}
\includegraphics[height=8cm,angle=-90]{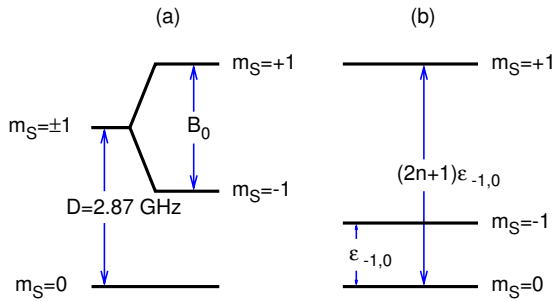}
\caption{\label{fig:levels}
(a) Energy levels of the electronic spin $S=1$ of a NV center. The states
$m_S=0$ and $m_S=-1$ form the relevant qubit subspace.
(b) The position of the energy levels when the high order harmonics in the control field
can induce the resonant leakage to the idle level $|m_S=1\rangle$.}
\end{figure}

The nitrogen-vacancy color center in diamond consists of a substitutional
nitrogen atom with an adjacent vacancy. Its many-electron orbital ground
state has a total spin \cite{Wrachtrup06} $S=1$. The combined effect of the
spin-orbit and spin-spin interactions leads to a single-axis spin anisotropy,
with the anisotropy axis directed along the $\langle 111\rangle$ crystallographic
direction (symmetry axis of the NV center) which we take as the $z$-axis.
As a result, the states $m_S=\pm 1$ and $m_S=0$ are split by $D=2.88$~GHz
(everywhere in this paper, we assume $\hbar=1$). The external static
field applied along the $z$-axis further splits the states $m_S=1$ and $m_S=-1$,
as shown schematically in Figure~\ref{fig:levels} (a),
and the total Hamiltonian of the system is
\begin{equation}
H_0 =DS^2_z+B_0S_z+B_x(t)S_x
\end{equation}
where $B_x(t)$ is the control field to be optimized.
When NV spin is used, for instance, as a qubit for quantum information processing,
only the two levels $m_S=0$ and $m_S=-1$ are employed.
The control field $B_x(t)$, whose amplitude is bounded $|B_x(t)|\leq M$,
is applied to drive the spin from the initial state $m_S=0$ to the final state $m_S=-1$.

Within the relevant subspace spanned by these two eigenstates,
the system is equivalent to a (pseudo)spin $s=1/2$.
The Hamiltonian for this spin, omitting the irrelevant energy shift, is
\begin{equation}
\label{eq:H2level}
H=Es^z+\sqrt 2 B_x(t)s^x
\end{equation}
where $E=(D+B_0)/2$, and $B_x(t)$ should steer the spin from
$|\downarrow\rangle$ to $|\uparrow\rangle$.
The optimal control for this problem
has been thoroughly investigated in Ref.~[\onlinecite{Boscain06}].
The time-optimal rotation of the (pseudo)spin 
is implemented by the bang-bang control, where
the control field $B_x(t)$ switches between $+M$ and $-M$.
The term {\sl bang} here, following the standard terminology, denotes
the interval between two neighboring switchings, i.e.\
during a bang the control field is constant.
When the maximal control field is strong, $M>\sqrt 2 E$,
the time-optimal rotation is achieved with exactly one switching\cite{Boscain06}
\begin{equation}
B_x(t) =
\begin{cases}
 M  &0\leq t<t_1\\
-M &t_1\leq t\leq T\\
\end{cases}
\end{equation}
where the switching time $t_1=[\pi\pm\arccos({2E^2}/{M^2})]/(2\sqrt{E^2+M^2/2})$,
and the optimal control time is
\begin{equation}
\label{eq:largeM}
T=\frac{\pi}{\sqrt{E^2+{M}^2/2}}\;.
\end{equation}
Corresponding spin trajectory on the Bloch sphere is rather simple:
the spin vector starts from the south pole
of the sphere ($|\downarrow\rangle$ state),
rotates about the axis $2E\hat z+\sqrt 2 M\hat x$ until it hits the equator,
then continues to rotate about the axis $2E\hat z-\sqrt 2 M\hat x$
all the way to the north pole ($|\uparrow\rangle$ state).
Here $\hat x$ and $\hat z$ are the unit vectors along $x$ and $z$ axes, respectively.
Note that the point $E=0$ is singular: a static
field of maximum amplitude would perform the necessary rotation twice faster than
the protocol of Ref.~\onlinecite{Boscain06}. However, this can be used
only in very special situation, in the nearest vicinity ($E\ll M$) of the point $E=0$,
so we do not consider this singular case below.

%
For weak driving $M<\sqrt 2E$, the duration of all bangs is the same,
except for the initial and the final bangs:
\begin{equation}
\label{eq:bangs}
\tilde B_x(t)=
(-1)^{k-1} M,~t\in[t_i+(k-1)t_m,t_i+k t_m)
\end{equation}
where $t_i$, $t_m$, and $t_f$ denote the durations of the initial,
middle, and the final bangs, respectively.
$n_\text{swt}$ is the number of times the field switchings between $+M$ and $-M$, and
$k=1,2,\cdots n_\text{swt}+1$ denotes the bangs.
There are also theorems relating $t_m$ and $t_f$ to $t_i$,
and limitations on the number of the field switchings.\cite{Boscain06}

Instead of the protocol of Ref.~\onlinecite{Boscain06},
a traditional approach for spin rotation employs a driving field along the $x$ axis
oscillating at Larmor frequency, $B_x(t)=M\cos(2Et)$.
This control works well in the weak-driving regime, $M\ll E$.
Within the rotating-wave approximation (RWA),\cite{Slichter} when
only the secular terms are retained in the control field,
the oscillating field is equivalent to a driving field with magnitude $M/2$
which rotates in the $x$--$y$ plane in step with the Larmor precession of the spin $s$.
The corresponding rotation time is $t_{\text{RWA}}=\sqrt 2 \pi/M$.
In a general situation, when the control field can be directed anywhere
in the $x$--$y$ plane, the rotating field is known to provide time-optimal
control. \cite{Boscain02} However, in reality it is very
difficult to create two perpendicular control fields with precisely locked directions,
amplitudes, and phases. Therefore, in the case of the oscillating driving along
the $x$ axis, half of the control field approximates the optimal two-axis rotating-wave
solution, while the other half (corresponding to the
omitted non-secular terms in the RWA) is wasted. The solution of
Ref.~\onlinecite{Boscain06} utilizes the full magnitude of the single-axis control in an optimal way.
At the same time, for $M\ll E$, its approximately periodic structure resembles the
oscillating driving, but the period (duration of the bangs) is slightly
different, and its time shape is rectangular rather than sinusoidal.

\begin{figure}
\includegraphics[height=8cm,angle=-90]{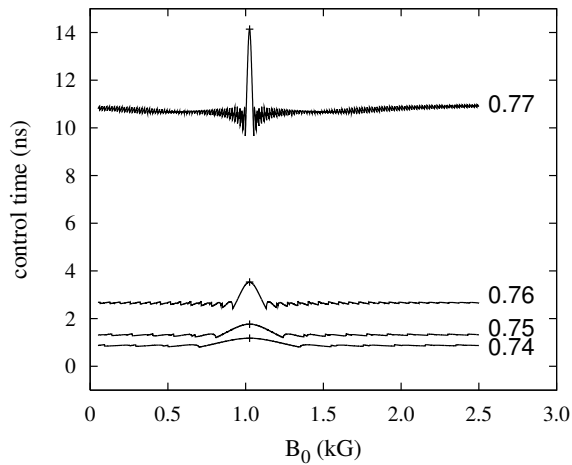}
\caption{\label{fig:topt} Spin rotation time as a function of the static field $B_0$ for
different amplitudes of the control field, $M=50,200,400,600$~MHz from top to bottom, respectively.
The rotation time at the peak of each curve is (accidentally) equal to $t_{\text{RWA}}$,
and is denoted by the points of different shapes.
The number on the right of each curve indicates the ratio of the control time
to $t_{\text{RWA}}$ at large bias $B_0$.}
\end{figure}


In a typical experiment, the maximum magnitude of the driving field $M$ is
fixed, limited by the setup details, while the static external field
$B_0$ is adjusted to provide the convenient conditions for driving the
rotation of the NV center spin from $m_S=0$ to $m_S=-1$.
Fig.~\ref{fig:topt}
shows the rotation time for the protocol of Ref.~\onlinecite{Boscain06}
as a function of the bias $B_0$ for different driving amplitudes
$M=200,~400,~600~$MHz. Overall, as expected, the rotation time decreases approximately as $1/M$.
At $B_0=1028$~G the distance $2E=D+B_0$ between
$m_S=0$ and $m_S=-1$ is zero (level crossing point),
and the rotation time is maximal.
The corresponding control time $2\pi/(\sqrt 2 M)$ is, accidentally, equal to $t_\text{RWA}$,
although RWA is not applicable here. Farther from the
level crossing point, the rotation time decreases. At very large distance $2E$
the rotation time almost saturates at the value
not very far from \cite{Boscain06} $t_{\text{RWA}}/\sqrt{2}$.
Therefore, the protocol of Ref.~\onlinecite{Boscain06} can decrease the rotation
time by about 25\% in comparison with the standard oscillatory $B_x(t)$.

\begin{figure}[htbp]
\includegraphics[height=8cm,angle=-90]{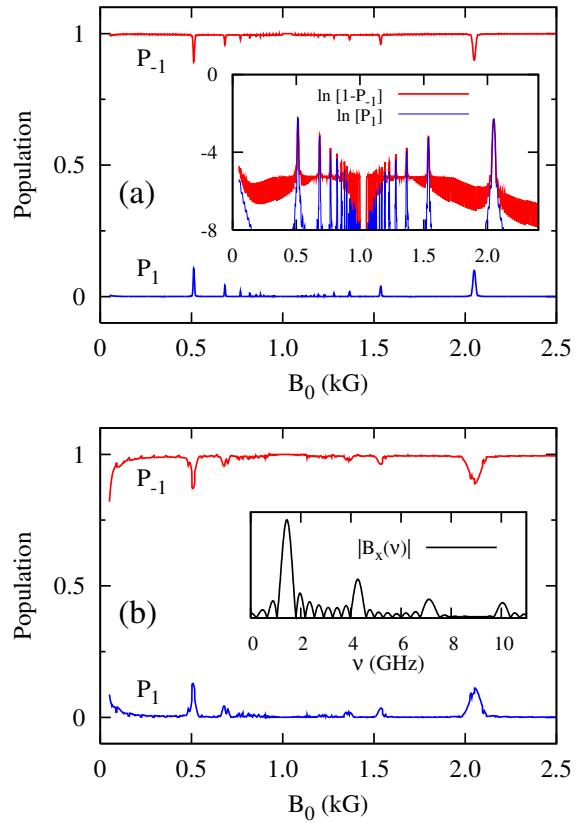}
\caption{\label{fig:Bp50} (Color online).
Dependence of the final population $P_{-1}$ (red line) and $P_1$ (blue line)
on the static field $B_0$,
after the application of the optimal control for the corresponding two-level system.
(a) The results for $M=50$~MHz. The inset shows the values $\ln[1-P_{-1}]$ (red)
and $\ln[P_1]$ (blue) vs.\ $B_0$. (b) The results for $M=200$~MHz. The
inset shows the Fourier transform of the bang-bang control, $B_x(\nu)$, for static bias $B_0=D/2$.
The first peak at $1.435$ GHz corresponds to the desired transition between levels
$m_S=0$ and $m_S=-1$,
while the second peak at $4.305$ GHz drives the unwanted transition to $m_S=1$ state.
}
\end{figure}

Fig.~\ref{fig:Bp50} shows the fidelity of the spin-1/2 protocol when applied
to a NV spin $S=1$. Except for a few sharp dips at certain bias fields, the final
population
$P_{-1}$ of the targeted level $m_S=-1$ stays high, and far from the level crossing
saturates at the values $0.999$, $0.994$, and $0.986$ for $M=50$, 200, and 600~MHz, respectively.
Thus, the protocol of Ref.~\onlinecite{Boscain06} provides good fidelity for $S=1$ spin
of NV center for most values of $B_0$. The loss of fidelity, including the
neighborhood of the dips, is predominantly due to leakage into the idle state $m_S=1$,
as seen in Fig.~\ref{fig:Bp50}, and the dips in $P_{-1}$ are accompanied by the peaks of $P_1$.

\begin{figure}[htbp]
\includegraphics[height=8cm,angle=-90]{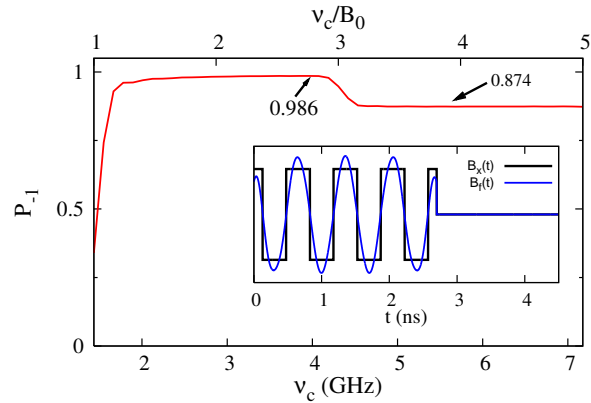}
\caption{\label{fig:cutoff} (Color online).
Final population $P_{-1}$ of the $m_S=-1$ state as a function of the cutoff
frequency $\nu_c$, for $M=200$ MHz and $B_0=D/2=1.435$ GHz
(where the energy gap between $m_S=0$ and $m_S=-1$ equals $B_0$).
When $\nu_c$ becomes smaller than $\epsilon_{+1,0}$, the leakage
to $m_S=+1$ is suppressed, and $P_{-1}$ abruptly increases.
The inset shows the time profile of the original bang-bang (black line)
and filtered (blue line) controls. The bang-bang control has
$t_i=t_f=0.1266$~ns, $t_m=0.3497$~ns, with $n_{\text{swt}}=8$ switchings.}
\end{figure}

Such leakage is particularly prominent when the energy levels satisfy the condition
\begin{eqnarray}
\label{eq:resonance}
\epsilon_{+1,0}=(2n+1)|\epsilon_{-1,0}|\;,
\end{eqnarray}
where $\epsilon_{\pm1,0}$ are the energy splitting between $m_S=\pm 1$ and $m_S=0$, respectively,
and $n=0,1,2\ldots$
In terms of $B_0$, these conditions are
\begin{eqnarray}
\label{eq:resonance2}
B_0=\frac{n}{n+1}D \text{ or } \frac{n+1}{n} D\;.
\end{eqnarray}
The increased leakage is determined by the (approximately) square-wave time profile of $B_x(t)$,
with (approximate) period $2t_m$, see Eq.~\ref{eq:bangs}. The Fourier expansion of the square wave contains
all frequencies $(2n+1)\pi/t_m$ with integer $n$, and these features are prominent in the
Fourier transform of the bang-bang control $B_x(t)$ (inset of Fig.~\ref{fig:Bp50}). The harmonics of the driving field
come in resonance with the transition between $m_S=0$ and $m_S=1$, see Fig.~\ref{fig:levels}, thus drastically
increasing the leakage. To demonstrate this, we set the external field at the
value $B_0=D/2$ ($n=1$) corresponding to the middle of the pronounced dip in Fig.~\ref{fig:Bp50}.
Instead of the actual bang-bang control $B_x(t)$, which leads to $P_{-1}=0.874$,
we used its filtered version $B_f(t)$, by cutting off the Fourier harmonics with frequencies
above certain value $\nu_c$. Also, to make the comparison fair,
the control time for filtered control was kept the same as the original
bang-bang control time (by turning off the filtered control $B_f(t)$ at longer times, see Fig.~\ref{fig:cutoff}).
Fig.~\ref{fig:cutoff} shows the dependence of $P_{-1}$
on the cutoff value: as soon as $\nu_c$ becomes smaller than $3D/2$, the value of $P_{-1}$
demonstrates a clear jump, since the unwanted resonant harmonic in the driving field is
filtered out almost completely, and the leakage to $m_S=1$ state is suppressed.

\section{Conclusions and discussions}

NV centers in diamond represent a very suitable system for fundamental studies
of quantum spin dynamics, and for many quantum spin-based applications.
In many applications so far, the spin of a NV center was used to represent
a qubit, with the states $m_S=0$ and $m_S=-1$ forming the relevant two-level subspace,
and the state $m_S=1$ remaining idle.
We investigated whether the spin $S=1$ of a NV center constitutes also a suitable
testbed for studying the time-optimal control protocol designed for two-level systems.
We found that by applying the optimal control protocol developed in Ref.~\onlinecite{Boscain06},
the rotation time from $m_S=0$ to $m_S=-1$ state can be reduced by $\sim 25$\% in
comparison with the standard sinusoidal oscillatory driving (within RWA approximation).
We checked how important is the unwanted leakage to the idle state
$m_S=1$. We found that this leakage, for most values of the external field, is negligible,
so that the NV center indeed could be a suitable platform for studying the time-optimal
control problems. We found, however, that near certain values of the bias field $B_0$,
the leakage to the idle state is drastically increased, and we demonstrated that
this is induced by higher-order harmonics in the bang-bang driving.
We show that by frequency filtering, this leakage can be noticeably suppressed.

Experimental implementation of the protocol studied above on NV centers seems
reasonable. Currently, the magnetic fields which drive the desired rotation
can be controlled on timescales of tens of picoseconds (or even faster with
specialized techniques). The driving field amplitudes of hundreds of MHz
have also been implemented in many experiments. \cite{deLange10,Fuchs09,Ryan10,Naydenov10}
Moreover, for faster rotation, it is important to mover far away from the
level crossing point at $B_0=1028$~G, so that negative external bias or very large external
field is more favorable for the rotation speedup.

It is important to note that the focus of this work is to check whether a NV spin
is a suitable testbed for experimental studies of the time-optimal protocols.
Although the protocol studied above does speed up the NV spin rotation, it
most probably presents a sub-optimal solution for the full system $S=1$.
However, the detailed studies of the time-optimal control of three-level systems
present a difficult mathematical problem, and are still lacking.
In this case, using an optimal solution within a restricted subspace
is a reasonable option, which may complement the numerical solution of the time-optimal
problem.

\acknowledgements
We thank D. D'Alessandro for numerous helpful discussions on time-optimal control
of spin systems and for very useful explanations. We also thank G. D. Fuchs and G. de Lange for many useful discussions
on experimental aspects of control of NV centers by time-varying magnetic fields.
Work at Ames Laboratory was supported by the Department of Energy --
Basic Energy Sciences under Contract No DE-AC02-07CH11358.

\end{document}